\pdfoutput=1

\documentclass[10pt,journal]{IEEEtran}
\ifCLASSOPTIONcompsoc
  \usepackage[nocompress]{cite}
\else
  \usepackage{cite}
\fi
\pdfoutput=1

\newtheorem{theorem}{Theorem}
\newtheorem{proposition}[theorem]{Proposition}
\newtheorem{lemma}[theorem]{Lemma}

\usepackage{graphicx}
\usepackage{subfig}
\usepackage{enumitem}
\usepackage{amsmath}
\usepackage{adjustbox}
\usepackage{amsfonts}
\usepackage{amssymb}

\usepackage{algorithm}
\usepackage{algpseudocode}

\begin{document}

\title{Towards the Methodology for Solving the\break
       Minimum Enclosing Ball and Related Problems}

\author{Michael~N.~Vrahatis
\IEEEcompsocitemizethanks{\IEEEcompsocthanksitem M.~N.~Vrahatis is with the Department
of Mathematics, University of Patras, GR-26110 Patras, Greece.\protect\\
E-mail: vrahatis@math.upatras.gr\,\, \&\,
vrahatis@upatras.gr}
}
\IEEEtitleabstractindextext{
\begin{abstract}
Methodology is provided towards the solution of the minimum enclosing ball problem.
This problem concerns the determination of the unique spherical surface of smallest radius enclosing a given bounded set
in the \emph{d}-dimensional Euclidean space.
Mathematical formulation and typical methods for solving this problem are presented.
Also, the paper is focused on areas that are related to this problem, namely:
(a) promise problems and property testing,
(b) theorems for partitioning and enclosing (covering) a set,
and (c) computation of the diameter of a set.
\end{abstract}

\thispagestyle{empty}

\begin{IEEEkeywords}
Formulation and methods for solving the MEB problem,
promise problems, property testing,
partitioning and enclosing (covering) theorems,
computing the diameter of a set.
\end{IEEEkeywords}}

\maketitle

\IEEEdisplaynontitleabstractindextext

\IEEEpeerreviewmaketitle

\ifCLASSOPTIONcompsoc
\IEEEraisesectionheading{\section{Introduction}\label{sec:introduction}}
\else
\section{Introduction and Mathematical Formulation}\label{sec:introduction}
\fi

\IEEEPARstart{M}{ethodology} is provided that is related to the solution of the \emph{minimum enclosing ball} (MEB) problem,
which refers to the determination of the unique spherical surface of smallest radius enclosing a given bounded subset of
the $d$-dimensional Euclidean space $\mathbb{R}^d$.
It can be considered that the earliest known statement of the MEB problem was first posed in \emph{circa}\/ 300 B.C.\
by Euclid\footnote{Euclid (\emph{c.}~323 -- \emph{c.} 285 B.C.), Greek mathematician, considered ``father of geometry''.}
in his seminal work \emph{``Elements''} and specifically in Book IV, Proposition 5
which refers to circumscribe a circle about a given triangle~\cite{Euclid1482}.

The MEB problem for finding the $d$-dimensional hypersphere (ball) with the smallest radius $r$ centered at the point $c$,
that encloses a set $P=\{p_1, p_2, \ldots, p_n\}$ of $n$ points in $\mathbb{R}^d$,
can be described as the following \emph{minimax optimization problem}\/ which involves both minimization over
the primal variable $c$ and maximization over the dual variable $i$:
\begin{equation}\label{eq:minmaxMEB}
 \mbox{MEB}(P): \kern0.3cm \min_c \max_i \| p_i - c\, \|, \kern1.2cm
\end{equation}
where $\| \cdot \|$ is the Euclidean norm in $\mathbb{R}^d$.
In the context of mathematical optimization the MEB problem can be formulated as a \emph{constrained optimization problem}\/
as follows::
 \begin{eqnarray}\label{eq:MEB}
\mbox{MEB}(P): \kern0.15cm \left\{
\begin{array}{l}
\displaystyle \min_{(c,\,r)}\,r , \\[0.2cm]
{\rm s.\,t.} \kern0.15cm  \| p_i - c \| \leqslant r ,  \kern0.3cm  p_i \in P.
\end{array}
\right.
\end{eqnarray}

A large variety of methods and algorithms for approximating the MEB and the diameter of a set have been proposed in the literature.
For mathematical approaches to the MEB and related to it problems, the interested reader is referred to
\cite{Sukharev1992,BoydV2004,Vrahatis2024a,Vrahatis2024b}
and the references therein.
In addition, the minimizers related to Problems (\ref{eq:minmaxMEB}) and (\ref{eq:MEB}) can be also computed by applying
\emph{intelligent optimization}\/ and \emph{computational intelligence}\/ methods,
including particle swarm optimization, differential evolution, memetic algorithms, among others.
These methods are capable of tackling non-differentiable, discontinuous, discrete, noisy and multimodal objective functions and in
general they have gained increasing popularity in recent years due to their relative simplicity and their ability to efficiently
and effectively handling several real-world applications
(see, \emph{e.g}.\
\cite{Dawkins1976,Schwefel1995,StornP1997,ClercK2002,LaskariPV2002a,LaskariPV2002b,ParsopoulosV2002a,ParsopoulosV2002b,Trelea2003,ParsopoulosV2010}).
Also, these methods have been widely used and applied for many years.
For example, for the cases at hand, the study of the performance of the particle swarm optimization method in coping with minimax and
constrained optimization problems has been initiated in 2002 (\emph{cf}.\ \cite{LaskariPV2002a} and \cite{ParsopoulosV2002b} respectively).
Related code can be found in \cite{ParsopoulosV2002a} and \cite{ParsopoulosV2010}.

It is worth mentioning that, the exploratory data analysis is an important first step in any data analysis
for identification of general patterns in a data set.
In real-world applications, these patterns include, among others, \emph{outliers}\/ of the data set that might be unexpected.
In such cases, the action of removing outliers during data preprocessing for enhancing the quality of the data set is important.
To this end, the \emph{minimum $k$-enclosing ball}\/ (M$k$EB) can be used for removing $n-k$ points
of the given $n$ points of a set $P$ in $\mathbb{R}^d$.
The M$k$EB problem is a generalization of the MEB one for finding a $k$-enclosing ball,
\emph{i.e}.\/ a ball with the smallest radius that encloses at least $k$ points of the set $P$.
It can be formulated as follows (\emph{cf}.~\cite{CavaleiroA2022}):
 \begin{eqnarray}\label{eq:MkEB}
\mbox{M}k\mbox{EB}(P): \kern0.0cm \left\{
\begin{array}{l}
\displaystyle \min_{(c,\,r)}\,r , \\[0.3cm]
{\rm s.\,t.} \kern0.15cm  \mbox{card}\bigl\{ p_i \in P : \|p_i - c \| \leqslant r\bigr\} \geqslant k, \\
\end{array}
\right.
\end{eqnarray}
where $\mbox{card}\{A\} \equiv |A|$ indicates the cardinality of a set $A$.
That is to say, the M$k$EB$(P)$ problem consists of finding a $k$-subset $X$ of $P$ (\emph{i.e}.\ $\mbox{card}\{X\}=k$)
whose minimum enclosing ball is one with the smallest radius among all $k$-enclosing balls.
Thus, the MEB$(P)$ Problem (\ref{eq:MEB}) is a particular case of the M$k$EB$(P)$ Problem (\ref{eq:MkEB}) for $k = n$.
It is worth noting that, in 2013
Shenma{\u\i}er\footnote{Vladimir V. Shenma{\u\i}er, Sobolev Institute of Mathematics, Russia.}
\cite{Shenmaier2013} proved that the M$k$EB problem is NP-hard in the strong sense \cite{Papadimitriou1994}
and obtain a \emph{polynomial-time approximation scheme}\/ (PTAS) for solving the problem with an arbitrary
relative error $\varepsilon$ in $O(n^{1/\varepsilon^2+1}d)$ time.

The general case of finding the smallest circle enclosing a given finite set of $n$ points in $\mathbb{R}^2$
was first given in 1857 by
Sylvester\footnote{James Joseph Sylvester (1814 -- 1897), English mathematician.}
\cite{Sylvester1857}.
In 1860 Sylvester gave for this problem a graphical solution method attributed to
Peirce\footnote{Charles Sanders Peirce (1839 -- 1914), American mathematician.}
\cite{Sylvester1860}.
The same method was independently given in 1885 by
Chrystal\footnote{George Chrystal (1851 -- 1911), Scottish mathematician.}
\cite{Chrystal1885}.
Thereafter, various methods for determining the MEB of a set of~$n$ points in the plane
have been proposed.

The first \emph{optimal linear-time method}\/ for solving the MEB problem for fixed dimension
has been presented in 1982 by
Megiddo\footnote{Nimrod Megiddo, Israeli mathematician and computer scientist.}
\cite{Megiddo1982}, (see also \cite{Megiddo1983}).
In addition, a simple and very fast in practice \emph{randomized algorithm}\/ to solve the problem,
also for $d=2$ and $d=3$, in expected linear time has been proposed in 1991 by
Welzl\footnote{Emmerich (Emo) Welzl (b.\ 1958), Austrian computer scientist.}
\cite{Welzl1991}.
A quite recent generalization of Welzl's algorithm for point sets lying in \emph{information-geometric spaces}\/
as well as an interesting application for solving a statistical model estimation problem by computing the
center of a finite set of univariate normal distributions, have been given in 2008 by
Nielsen\footnote{Frank Nielsen, PhD in 1996 in computational geometry from the INRIA Sophia-Antipolis, Universit{\'e} de Nice, France.}
and
Nock\footnote{Richard Nock, PhD in 1998 and HDR in 2002 in computer science from the University of Montpellier, France.}
\cite{NielsenN2008}.

The paper at hand aims to contribute towards the methodology for tackling the MEB problem and
the similar and related to it issues and aspects.

\section{Typical Methods for Solving the MEB Problem}\label{sec:RelAppl}

\noindent
\textbf{Elzinga-Hearn method:}\/
In 1972
Elzinga\footnote{Donald Jack Elzinga (b. 1939), American chemical engineer and mathematician.}
and
Hearn\footnote{Donald William Hearn (1939 -- 2021), American mathematician.}
\cite{ElzingaH1972}
presented a method for tackling the MEB problem for a set $P=\{p_1, p_2, \ldots, p_n\}$
of $n$ points in $\mathbb{R}^d$.
They formulated the MEB problem as a \emph{convex programming}\/ problem and
gave a \emph{finite decomposition}\/ algorithm based
on the \emph{simplex method}\/ of \emph{quadratic programming}.
Specifically, by representing the given points $p_i$, $i=1,2,\ldots,n$ as column vectors
and assuming that the scalar $s$ is the square of the radius $r$ of a sphere centered at the point $c$,
they formulated in~\cite{ElzingaH1972} the MEB Problem (\ref{eq:MEB}) to the following equivalent convex programming problem:
\begin{eqnarray}\label{eq:MEB-PP}
\mbox{MEB}_{\rm pp}(P): \kern0.15cm \left\{
\begin{array}{l}
\displaystyle \min_{(c,\,s)}\,s , \\[0.3cm]
{\rm s.\,t.} \kern0.15cm  s \geqslant (p_i - c)^{\top}
(p_i - c) ,  \kern0.3cm  p_i \in P .
\end{array}
\right.
\end{eqnarray}

The above Problem (\ref{eq:MEB-PP}) is referred to as the \emph{primal problem}\/ (PP)
which constitutes the original optimization problem.

The
Kuhn\footnote{Harold William Kuhn (1925 -- 2014), American mathematician.}
and
Tucker\footnote{Albert William Tucker (1905 -- 1995), Canadian mathematician.}
conditions (\emph{cf}.\ \cite{KuhnT1951}), that are necessary and sufficient for Problem (\ref{eq:MEB-PP}),
guarantee the existence of \emph{multipliers} $\lambda^*_i$ for $i = 1, 2, \ldots, n$, such that:
\begin{eqnarray}\label{eq:KTC}
\hfill
\left\{
\begin{array}{l}
\displaystyle \sum_{i=1}^n \lambda^*_i =1,\\[0.5cm]
\displaystyle\sum_{i=1}^n \lambda^*_i (p_i - c^*) = 0,\\[0.6cm]
\lambda^*_i \bigl( s^* - (p_i - c^*)^{\top} (p_i - c^*)\bigr) =0, \kern0.17cm  i = 1, 2, \ldots, n,\\[0.4cm]
\lambda^*_i \geqslant 0, \kern0.3cm  i = 1, 2, \ldots, n,\\[0.3cm]
 s^* - (p_i - c^*)^{\top} (p_i - c^*) \geqslant 0, \kern0.3cm  i = 1, 2, \ldots, n.
\end{array}
\right.
\end{eqnarray}

The \emph{Kuhn-Tucker Conditions}\/ (\ref{eq:KTC}) indicate that the center $c^*$ of the MEB
with $s^*$ the square of its radius will be a convex combination of the points which
lie on the surface of the MEB.
This is so because it holds $\lambda^*_i = 0$ for any point $p_i$ that lie in its interior.

It is worth mentioning here that, in general, the \emph{convex hull}, ${\rm conv}\,\!P$, of a set $P$ in $\mathbb{R}^d$
can be obtained by carrying out all convex combinations of points of $P$ (\emph{cf}.\ \cite{Rockafellar1970}).
On the other hand, due to a theorem of
Carath\'{e}odory\footnote{Constantin Carath\'{e}odory (1873 -- 1950), Greek mathematician.}
(see Theorem \ref{CaratheodoryTh} below)
it is not necessary to perform combinations involving more than $d + 1$ elements at a time.
Therefore, convex combinations of vectors~$x_i$ of the form $\sum_{i=1}^{m} \alpha_i x_i$
(where the real numbers $\alpha_i$ satisfy $\alpha_i \geqslant 0$ and $\sum_{i=1}^{m} \alpha_i = 1$)
can be performed where $m \leqslant d +1$ or $m=d + 1$ in the case of non-distinct vectors~$x_i$.

The above mentioned Carath\'{e}odory's theorem in convexity theory states that:
\begin{theorem}[Carath\'{e}odory's theorem (1907)~\cite{Caratheodory1907}]\label{CaratheodoryTh}
Any point in the convex hull of a finite point set in $\mathbb{R}^d$ is a convex combination
of some at most $d + 1$ of these points.
\end{theorem}

This important theorem in convex geometry has been given in 1907~\cite{Caratheodory1907}.
It is worth noting that, in 1970
Rockafellar\footnote{Ralph Tyrrell Rockafellar (b.\ 1935), American mathematician.}
\cite{Rockafellar1970} pointed out that:
\emph{``Carath\'{e}odory's theorem is the fundamental dimensionality result in convexity theory,
and it is the source of many other results in which dimensionality is prominent.''}
Based on Carath\'{e}odory's Theorem \ref{CaratheodoryTh} the following result has been presented by
Elzinga and Hearn in \cite{ElzingaH1972}:
\begin{lemma}[Elzinga-Hearn lemma (1972)~\cite{ElzingaH1972}]\label{ElzingaLem-1}
The center $c^*$ of the MEB can be expressed as a convex combination of at most $d + 1$ of the given points.
\end{lemma}

In addition, they pointed out that the existence of the MEB $(s^*, c^*)$, where $s^*$ is the square of its radius
and $c^*$ is its center, is evident. Also, since the minimand of
\begin{equation}\label{eq:minimand}
\min_{(s,\, c)} \max_i\,\, (p_i - c)^{\top} (p_i - c),
\end{equation}
is strictly convex, then the following result holds:
\begin{lemma}[Elzinga-Hearn lemma (1972)~\cite{ElzingaH1972}]\label{ElzingaLem-2}
The MEB
$(s^*, c^*)$ exists and is unique.
\end{lemma}

In 1961
Wolfe\footnote{Philip Starr Wolfe (1927 -- 2016), American mathematician.}
\cite{Wolfe1961}
pointed out that, in general, in the \emph{duality theory}\/ for mathematical programming
a \emph{duality theorem}\/ is the statement of a relationship of a certain kind
between two mathematical programming problems that has three characteristics.
Specifically:
\begin{itemize}
\leftskip0.05cm
\itemsep=3pt
\item[(a)]
one problem, named the \emph{primal problem}~(PP) concerns to a constrained minimization problem,
and the second one, named the \emph{dual problem}~(DP) concerns to a constrained maximization problem,
\item[(b)]the existence of a solution to one of these
problems ensures the existence of a solution to the other problem,
therefore their respective extreme values are equal, and
\item[(c)] in the case where the constraints of one problem are consistent, while
those of the other are not, there is a sequence of points satisfying the constraints of the
first on which its objective function tends to infinity.
\end{itemize}

The \emph{Wolfe's dual problem}\/ ($\mbox{W}_{\rm dp}$) for the primal problem is
the following (\emph{cf}.\ \cite{Wolfe1961,ElzingaH1972})\/:
\begin{eqnarray}\label{eq:WD}
\mbox{W}_{\rm dp}(P):
\left\{
\begin{array}{ll}
\displaystyle \!\!\!\max_{(s,\, c,\, \lambda)}\kern-0.8cm&\!\!  s  +  \displaystyle \sum_{i=1}^n \lambda_i
\bigl((p_i - c)^{\top} (p_i - c) - s\bigr),\\[0.5cm]
{\rm s.\,t.} &\!\!\displaystyle \sum_{i=1}^n \lambda_i =1, \\[0.5cm]
&\!\!\displaystyle\sum_{i=1}^n \lambda_i (p_i - c) = 0, \\[0.5cm]
&\!\!\lambda_i \geqslant 0, \kern0.3cm  i = 1, 2, \ldots, n . \\
\end{array}
\right.
\end{eqnarray}

It is worth noting that, in general, \emph{converse duality}\/ does not hold in the sense that
the solution of the dual problem provides the solution of the primal one.
However, in the case of the MEB problem converse duality holds.
Specifically, in 1962
Huard\footnote{Pierre (Huard de la Marre) Huard (1927 -- 2022), French mathematician.}
\cite{Huard1962} has shown converse duality for certain ``partially linear'' problems,
of which Problem (\ref{eq:MEB-PP}) is a special case (\emph{cf}.\ \cite{ElzingaH1972}).

An equivalent to Wolfe's dual problem has been proposed in \cite{ElzingaH1972}
where the authors gave the following dual problem named
\emph{quadratic programming dual problem}\/ ($\mbox{QP}_{\rm dp}$)\/:
\begin{eqnarray}\label{eq:QPD}
\mbox{QP}_{\rm dp}(P):
\left\{
\begin{array}{ll}
\displaystyle \!\max_\lambda \kern-1.2cm&\!\! \displaystyle \sum_{i=1}^n
\lambda_i \bigl(p_i^{\top} p_i\bigr) - \lambda^{\top} \bigl(A^{\top}\! A\bigr) \lambda
,\\[0.5cm]
{\rm s.\,t.} &\!\!\displaystyle \sum_{i=1}^n \lambda_i =1, \\[0.5cm]
&\!\!\lambda_i \geqslant 0, \kern0.3cm  i = 1, 2, \ldots, n , \\
\end{array}
\right.
\end{eqnarray}
where $\lambda = (\lambda_1, \lambda_2, \ldots, \lambda_n)^{\top}$ and the matrix $A \in \mathbb{R}^{d\times n}$
has columns the $p_i$ for $i = 1, 2, \ldots, n$.
Based on the above, the authors of \cite{ElzingaH1972}
provided the following theorems\/:
\begin{theorem}[Elzinga-Hearn theorem (1972)~\cite{ElzingaH1972}\/]\label{ElzingaThm-1}
Wolfe's dual Problem (\ref{eq:WD}) is equivalent to the concave quadratic programming dual Problem (\ref{eq:QPD}),
with
\begin{equation}\label{eq:c}
c = \sum_{i=1}^n \lambda_i\/p_i ,
\end{equation}
and
\begin{equation}\label{eq:s}
s = \sum_{i=1}^n \lambda_i (p_i - c)^{\top} (p_i - c).
\end{equation}
\end{theorem}
\begin{theorem}[Elzinga-Hearn theorem (1972)~\cite{ElzingaH1972}\/]\label{ElzingaThm-2}
Assume that $(s^*, c^*, \lambda^*)$ solves the concave quadratic programming
dual Problem (\ref{eq:QPD}) then $(s^*, c^*)$ solves the primal Problem (\ref{eq:MEB-PP}).
\end{theorem}

By Theorem \ref{ElzingaThm-2} it is evident that the converse duality holds.
On the other hand, the solution of the corresponding quadratic problem requires a very large amount of computer storage.
To this end, the authors developed a \emph{finite decomposition algorithm}, based on the \emph{simplex method}\/
of quadratic programming, that makes the computer storage requirements independent of the number of points and
computing time \emph{approximately linear}\/ in the number of points.

\medskip
\noindent
\textbf{Hopp-Reeve method:}\/
In 1996
Hopp\footnote{Theodore H. Hopp, American computer scientist at National Institute of Standards and Technology (NIST), USA.}
and
Reeve\footnote{Charles P. Reeve, American computer scientist at National Institute of Standards and Technology (NIST), USA.}
\cite{HoppR1996} proposed a simple and efficient method for computing the MEB
of a set of $n$ points $P=\{p_1, p_2, \ldots, p_n\} \subset \mathbb{R}^d$.
Their method geometrically constructs the MEB using an iterated two-step procedure.
At the beginning of each iteration, a non-empty set $Q\subseteq P$ exists such that all points in $Q$ are affinely independent.
Furthermore, an \emph{enclosing ball}\/ (EB) for $P$ exists such that each point in $Q$ lies on the surface of the EB.
Note that the EB is not necessarily the MEB for $Q$.
Assume that $c$ is the center of the EB which initially may be taken as point $p_1$ and the set $Q$ as the point farthest from $p_1$.
The points in $Q$ before each iteration are called \emph{candidate points}, while the points in $Q$ after the final iteration are
called \emph{constraining points}.
A high level description of the iterated steps in constructing the MEB are exhibited in Algorithm~\ref{algo:HR}.
\begin{algorithm}[htb]
\textbf{Input:}
The sets $P=\{p_1, p_2, \ldots, p_n\} \subset \mathbb{R}^d$ and $Q\subseteq P$.
\begin{algorithmic}[1]
\State Compute the center $t$ of the MEB for $Q$, then remove from $Q$ any points which do not constrain the
MEB. The point $t$ then serves as a target.
\State Shrink the EB by moving the center $c$ toward $t$ while maintaining all points in $Q$ on the sphere surface.
If the surface of the shrinking EB contacts a point in $P$ but not in $Q$, fix $c$ and add the new point to $Q$.
\end{algorithmic}
\textbf{Output:}
The MEB for $P$ has been found when either:
\begin{itemize}
\leftskip0.0cm
\item[(a)]
$Q$ contains exactly $d+1$ points at the end of Step~1 (in this case $c$ and $t$ will coincide at the start of Step~2), or
\item[(b)]
the center $c$ of the shrinking EB in Step~2 reaches target $t$ without the sphere surface contacting a new point.
\end{itemize}
\caption{Hopp and Reeve (1996) \cite{HoppR1996}:\\
\emph{Geometrically construction of the MEB.}}\label{algo:HR}
\end{algorithm}

The \emph{expected computing time}\/ of the Hopp-Reeve algorithm, in the \emph{worst case}\/
with all the points near the ball surface, was estimated at $O(n d^{\,2.3})$.
The authors pointed out that the number of points $n$ and dimension $d$ are bounded above only by computer storage limitations.
Furthermore, in infinite-precision arithmetic the MEB must always be found in at most
\begin{equation}\label{eq:HR-N}
N_{\rm iter} = \sum_{i=2}^{\min \{n,\, d+1\}} \frac{n !}{i !\, (n-i)!}
\end{equation}
iterations. However, in finite-precision arithmetic an endless loop is likely to occur among several sets of points.
Thus, to cope with this possibility, in the implementation of this algorithm should be incorporated
an upper bound on the number of iterations.

\medskip
\noindent
\textbf{B\u{a}doiu-Clarkson method:}\/
In 2003
B\u{a}doiu\footnote{Mihai B\u{a}doiu, PhD in 2006 in computer science from the Massachusetts Institute of Technology, USA.}
and
Clarkson\footnote{Kenneth Lee Clarkson, PhD in 1984 in computer science from the Stanford University, USA.}
\cite{BadoiuC2003} proposed a simple iterative approximation algorithm for tackling the MEB problem
for a set $P=\{p_1, p_2, \ldots, p_n\}$ of~$n$ points in $\mathbb{R}^d$ (\emph{cf}.\ \cite{Shenmaier2013}).
This algorithm is exhibited in Algorithm~\ref{algo:MEB}.
\begin{algorithm}[htb]
\textbf{Input:}
The set $P=\{p_1, p_2, \ldots, p_n\} \subset \mathbb{R}^d$ and $k \in [1,n]$.
\begin{algorithmic}
\State Choose an arbitrary point $p_j \in P$, $(1\leqslant j \leqslant n)$.
\State Set $c_1 = p_j$.
		\For {$i=2 \mbox{\, {\bf to}\, } k$}
\State Find the point $p_i \in P$ farthest away from $c_{i-1}$.
\State Set $c_{i} = c_{i-1} + (p_i -c_{i-1})/i$.
		\EndFor
\State Set $r_k = \max_{x\in P} \| x - c_{k}\|$.
\end{algorithmic}
\textbf{Output:}
The ball of radius $r_k$ centered at the point $c_{k}$.
\caption{B\u{a}doiu and Clarkson (2003) \cite{BadoiuC2003}:\\
\emph{Approximation of the MEB problem.}}\label{algo:MEB}
\end{algorithm}

Furthermore, the authors gave the following result that is related to Algorithm~\ref{algo:MEB}.
\begin{proposition}[B\u{a}doiu-Clarkson proposition (2003) \cite{BadoiuC2003}]
 If $c$ and $r$  are the center and radius of a minimal ball enclosing the set $P$ then
 $\| c - c_i\| \leqslant r / \sqrt{i}$ for all $i$.
\end{proposition}

\section{Promise Problems, Property Testing and Partitioning Theorems}\label{sec:PropTest}

The MEB problem can be tackled by applying enclosing and partitioning theorems and it is, among others,
 one of the most fundamental issues in clustering.
Specifically, \emph{clustering}\/ with respect to the \emph{diameter}\/ and the \emph{radius} costs,
is the task of \emph{partitioning}\/ a set of points in $\mathbb{R}^d$ to subsets, where items in the same subset,
named \emph{cluster}, are similar to each other, compared to items in other clusters
(see, \emph{e.g.} \cite{AlonDPR2003,AlonDPR2004}).
Usually, clustering problems arise in the analysis of large data sets.
Thus, the \emph{approximate clustering via core-sets} is used for clustering of a set of points
in $\mathbb{R}^d$ (for large $d$\,) by extracting properly a small set of points named \emph{core-set}\/
that ``represents'' the given set of points~\cite{BadoiuHI2002}.
These issues lie within the domains of promise problems and property testing.

\medskip
\noindent
\textbf{Promise problems:}\/
The notion of the \emph{promise problem} has been introduced and studied in 1984 by
Even\footnote{Shimon Even (1935 -- 2004), Israeli computer scientist.},
Selman\footnote{Alan Louis Selman (1941 -- 2021), American mathematician and computer scientist.}
and Yacobi\footnote{Yacov Yacobi, PhD in 1980 in electrical engineering from the Technion Israel Institute of Technology.}
\cite{EvenSY1984} (\emph{cf}.\ \cite{Goldreich2006a}).
In general, a promise problem can be considered as a formulation of a partial decision problem and the complexity
issues related to it have been arisen from considerations about cracking problems for \emph{public-key cryptosystems}\/
(\emph{cf}.\ \cite{EvenSY1984,EvenY1980}).
The authors of \cite{EvenSY1984} considered a \emph{decision problem}\/ as a \emph{predicate}\/ $P(x)$.
They pointed out that the objective is the determination whether there exists an algorithm $\mathcal{A}$
that solves the problem, \emph{i.e}.\ such that $\mathcal{A}(x)$ converges for all input instances $x$ and
\[
\forall\,x\,\bigl[\mathcal{A}(x) = \texttt{YES} \longleftrightarrow P(x)\bigr].
\]
Also, they pointed out that, in general, there are problems for which only a subclass of the domain of all instances is required.
The authors named these problems \emph{promise problems}\/ and suggested that a promise problem has the structure:
\begin{itemize}
\leftskip0.00cm
\item[(a)]
\emph{input}\/ $x$,
\item[(b)]
\emph{promise}\/ $Q(x$),
\item[(c)]
\emph{property}\/ $R(x)$,
\end{itemize}
where $Q$ and $R$ are predicates. Furthermore, they deduced that, formally:
\begin{itemize}
\leftskip-0.00cm
\item[(a)]
A \emph{promise problem}\/ is a pair of \emph{predicates}\/ $(Q, R)$.
\item[(b)]
The predicate $Q$ is called the \emph{promise}.
\item[(c)]
A \emph{deterministic
Turing\footnote{Alan Mathison Turing (1912 -- 1954), English mathematician and
computer scientist.}
machine}\/ $M$ solves the promise problem $(Q, R)$ if
\[
\forall\,x\,\Bigl[Q(x)\longrightarrow \bigl[ M(x)\!\downarrow\,
\wedge\, \bigl(M(x) = \texttt{YES}
\longleftrightarrow R(x)\bigr)\bigr]\Bigr],
\]
where the notation $M(x)\!\downarrow$\, indicates that $M$ eventually halts on input $x$.
\item[(d)]
A promise problem $(Q, R)$ is \emph{solvable}\/ if there exists a Turing machine $M$ that solves it.
\item[(e)]
If a Turing machine $M$ solves $(Q, R)$, then the language $L(M)$ accepted by $M$ is a \emph{solution} to $(Q, R)$.
\end{itemize}

It is worth mentioning that, in 2006
Goldreich\footnote{Oded Goldreich (b.\ 1957), Israeli computer scientist.}
\cite{Goldreich2006a}
pointed out that any decision problem is a promise problem, although in some cases the promise is trivial or tractable.
He noticed that the promise problems constitute a natural generalization of the decision problems and, in many cases, a promise
problem provides the more natural formulation of a decision problem (\emph{cf}.\ \cite{Goldreich2006b}).
Also, the author considered that, formally, a promise problem refers to a three-way partition of the set of all strings into
three subsets; namely:
\begin{itemize}
\leftskip-0.00cm
\item[(a)]
The set of strings representing \texttt{YES}-\emph{instances}.
\item[(b)]
The set of strings representing \texttt{NO}-\emph{instances}.
\item[(c)]
The set of \emph{disallowed}\/ strings that represent neither \texttt{YES}-instances nor \texttt{NO}-instances.
\end{itemize}
The union of the set of all the \texttt{YES}-instances and the set of all the \texttt{NO}-instances is called the \emph{promise}.
The set of instances that satisfy the promise is named \emph{promise set}, while the set of the rest instances is called
the set of instances that \emph{violate the promise}.
Hence, an eventual \emph{decider}\/ (\emph{e.g}.\ algorithm or procedure) for \emph{solving the promise problem}\/
has only to distinguish \texttt{YES}-instances from \texttt{NO}-instances,
while, for the inputs that violate the promise is allowed arbitrary behavior.
Consequently, is only required to detect that an input is either a \texttt{YES}-instance or a \texttt{NO}-instance.

In addition, the author of~\cite{Goldreich2006b} has commented that the discrepancy between the formulation of
intuitive promise problem and the standard formulation of decision problems can be easily tackled in the case where
there exists an efficient algorithm for determining \emph{membership}\/ in the promise set.
In this instance, the promise problem is \emph{computationally equivalent}\/ to deciding
membership in the set of \texttt{YES}-instances. On the other hand, in the case where the promise set
is not tractable, the terminology of promise problems is necessary (\emph{cf}.\ \cite{Goldreich2006b}).
For further details on promise problems and their applications, the interested reader is referred to \emph{e.g}.\
\cite{EvenSY1984,Goldreich2006a,EvenY1980,Goldreich2006b,GoncalvesKSY2023}.

In 2023
Gon\c{c}alves\footnote{Felipe Gon\c{c}alves, PhD in 2016 in mathematics from the IMPA--Rio de Janeiro, Brazil.},
Keren\footnote{Daniel Keren, PhD in 1991 in computer science from the Hebrew University of Jerusalem, Israel.},
Shahar\footnote{Amit Shahar, Department of Computer Science, University of Haifa, Israel.}
and
Yehuda\footnote{Gal Yehuda, Department of Computer Science, Technion-IIT, Israel.}
\cite{GoncalvesKSY2023}
proposed a probabilistic approach for the \emph{clustering promise problem}.
They pointed out that, their main contribution is to bring together ideas from the \emph{theory of random fields}\/
and \emph{streaming algorithms}\/ and their goal was to devise algorithms for answering the clustering promise problem
quickly, while having full access to the dataset. They defined the clustering promise problem:
\[
\kern-0.15cm
\left\{
\begin{array}{ll}
\Pi = \Pi_{\texttt{YES}}\, \cup\, \Pi_{\texttt{NO}},\\[0.1cm]
\Pi_{\texttt{YES}} = \bigl\{P : P \mbox{ is } (k_1, \varepsilon) \mbox{-clusterable}\bigr\},\\[0.1cm]
\Pi_{\texttt{NO}} = \bigl\{P : P \mbox{ is } (k_2, \delta) \mbox{-far from being clusterable}\bigr\},
\end{array}
\right.
\]
where, given the parameters $k_1, k_2 \in \mathbb{N}$ and $\varepsilon, \delta \in \mathbb{R}_+$,
a set of vectors $P \subseteq \mathbb{R}^d$ is defined to be $(k_1, \varepsilon)$-\emph{clusterable}\/
if there are $k_1$ balls of radius $\varepsilon$ that cover $P$;
while $P \subseteq \mathbb{R}^d$ is defined to be $(k_2, \delta)$-\emph{far from being clusterable}\/
if there are at least $k_2$ vectors in $P$, with all pairwise distances at least $\delta$.
For realizing the relation between the conditions $\Pi_{\texttt{YES}}$ and $\Pi_{\texttt{NO}}$
the authors of \cite{GoncalvesKSY2023} gave the following result:
\begin{lemma}[Gon\c{c}alves-Keren-Shahar-Yehuda lemma (2023) \cite{GoncalvesKSY2023}]
If the set of vectors $P \subseteq \mathbb{R}^d$  has at most $\ell$ vectors
with pairwise distance at least $\delta$, then its minimal cover with balls of radius $\varepsilon$ is of size
at most $\ell c(d) \bigl( \delta / \varepsilon\bigr)^d$, where $c(d)$ is a function which depends only on $d$.
\end{lemma}

Also, they provided a probabilistic algorithm for distinguishing the two issues of $\Pi_{\texttt{YES}}$
and $\Pi_{\texttt{NO}}$.
Their algorithm determines a decision using only the extreme values of a scalar valued \emph{hash function},
defined by a \emph{random field}\/ on the set $P$ and is particularly suitable for distributed and
online settings (\emph{cf}.\ \cite{GoncalvesKSY2023}).
It is worth recalling that, in general, a hash function is any function that can be used to map data
of arbitrary size to fixed-size values;
while a random field over $\mathbb{R}^d$ can be considered as a function that maps each $x \in \mathbb{R}^d$
to a random variable $f(x)$ over $\mathbb{R}$.
The authors noted that, the hash function can be defined using a random field as well as
they explicitly shown how such random fields can be constructed (\emph{cf}.\ \cite{GoncalvesKSY2023}).

\medskip
\noindent
\textbf{Property testing:}\/
The \emph{property testing}\/ has been primarily defined in 1996 by
Rubinfeld\footnote{Ronitt Rubinfeld (b. 1964), American electrical engineer and computer scientist.}
and
Sudan\footnote{Madhu Sudan (b. 1966), Indian-American computer scientist.}
\cite{RubinfeldS1996}
in the context of \emph{program testing}, \emph{cf}.\ \cite{AlonDPR2004},
(see also, \emph{e.g}.\ \cite{BlumK1989,BlumLR1993,BlumK1995}).
The authors pointed out that, the study of \emph{program checkers}, \emph{self-testing programs} and
\emph{self-correcting programs} has been introduced in order to allow one to use a program to compute
a function without trusting that the program works correctly.
This issue leads to the search for \emph{robust characterizations of functions}.
The authors rendered this notion precise and provided such a characterization for polynomials.
They considered characterizations of multivariate polynomials over various domains,
\emph{e.g}.\ rings of the form $\mathbb{Z}_m$ or finite fields.

Also, they introduced the concept of the \emph{robust characterization of a family of functions}\/
$\mathcal{F}$ with respect to a set of neighborhoods $\mathcal{N}$, where $\mathcal{N}$ consists of a set
of tuples in the domain of the functions in $\mathcal{F}$. The authors defined that:
``A property $\mathcal{P}$ over a collection of neighborhoods $\mathcal{N}$ is
an $(\varepsilon, \delta)$-\emph{robust characterization of}\/ $\mathcal{F}$,
if whenever a function satisfies $\mathcal{P}$ on all but a $\delta$-\emph{fraction}\/ of the
neighborhoods in $\mathcal{N}$, it is $\varepsilon$-\emph{close}\/ to some function $f \in \mathcal{F}$.
Moreover, all the members of $\mathcal{F}$ satisfy $\mathcal{P}$ on all neighborhoods in $\mathcal{N}$.''

It is worth noting that, from this characterization, the authors obtained applications related to
the construction of simple and efficient \emph{self-testers}\/ for polynomial functions.
Furthermore, the characterizations provided results in the area of \emph{coding theory}\/ by giving
extremely fast and efficient \emph{error-detecting schemata}\/ for some well-known codes.
In addition, the authors highlighted that, these error-detection schemata play
an important role in subsequent results related to the hardness of approximating various NP-optimization problems.

In brief, the property testing can be regarded as the testing of a global property with local inspection.
It is worth noting that, in 2008
Ron\footnote{Dana Ron (b. 1964), Israeli computer scientist.}
\cite{Ron2008} (\emph{cf}.\ \cite{RubinfeldS1996,GoldreichGR1998})
pointed out that:
\emph{``Property testing is the study of the following class of problems:
Given the ability to perform (local) queries concerning a particular object the problem is to determine whether the
object has a predetermined (global) property or differs significantly from any object that has the property.
In the latter case we say it is far from (having) the property. The algorithm is allowed a small probability of failure,
and typically it inspects only a small part of the whole object.''}
Also, the author presented as example the following two cases:
(a) the object may be a graph and the property is that it is bipartite, and
(b) the object may be a function and the property is that it is linear.

It is worth recalling and emphasising that, the concept of the property testing has been activated
for designing and development of \emph{super-fast}\/ algorithms for analysing the
\emph{global structure}\/ of datasets that are too large to read in their entirety in a reasonable time.
These algorithms have direct access to items of a \emph{huge}\/ input data set and determine whether
this data set satisfies a desired \emph{predetermined global property}, or is \emph{far}\/ from satisfying it.
Also, these algorithms manage relatively small portions of the data set and their complexity is measured
in terms of the number of accesses to the input.
The algorithms that are used are necessarily \emph{randomized}, otherwise they may be drawing a conclusion
from an atypical portion of the input.
In addition, they are \emph{approximate}, because it is not expected the algorithm to produce an exact
answer having examined only a portion of the input.
For further details regarding issues and aspects of the property testing,
the interested reader is referred to \emph{e.g}.\
\cite{AlonDPR2003,AlonDPR2004,Ron2008,CzumajSZ2000,CzumajS2001,Goldreich2017,ChakrabortyPRS2018,Ron2009,% BREAK FOR THE CITE CONTINUATION
Goldreich2010,GoldreichW2022,GoldreichR2023,NewmanV2024}.

It is worth mentioning that, in 2003
Alon\footnote{Noga Alon (b. 1956), Israeli mathematician.},
Dar\footnote{Seannie Dar, PhD in 1997 in convex and discrete geometry from the Tel Aviv University, Israel.},
Parnas\footnote{Michal Parnas, PhD in 1994 in computer science from the Hebrew University of Jerusalem, Israel.}
and
Ron\footnotemark[33]
\cite{AlonDPR2003} (\emph{cf}.\ \cite{AlonDPR2004}) investigated the clustering problem in the context of property testing.
Specifically, the authors discussed the \emph{testing of clustering}\/ in the framework of property testing.
They defined that, a data set $P$ of $n$ points equipped with some metric is $(k, b)$-\emph{clusterable}\/ if it can be partitioned
into $k$ sets with cost at most $b$.
The authors considered as \emph{cost}\/
(a) the \emph{radius}, that is the minimum number $r$ such that each cluster can be enclosed in a ball of radius $r$, or
(b) the \emph{diameter}, that is the maximal distance between any two points in a cluster
(\emph{cf}.\ \cite{CzumajS2001}).
They noted that, the data set is $\varepsilon$-\emph{far}\/ from being $\bigl(k, (1+ \beta)b\bigr)$-\emph{clusterable}\/
if at least $\varepsilon n$ points must be removed from $P$ so that the remaining data is $\bigl(k, (1+ \beta)b\bigr)$-clusterable.
Also, the authors described and analyzed algorithms that use a sample of size polynomial in $k$ and
$1/\varepsilon$ and independent of ${\rm card}\{P\}=n$.
They noted that, the dependence on $\beta$ and on the dimension, $d$, of the points varies with the different algorithms.
In addition, they pointed out that, this kind of algorithms are particular useful in the case where the set
of points $P$ is very large and it may not even be feasible to read all of it.
For general metrics, the authors proposed an algorithm for the case $\beta = 1$, that requires a
sample of size $O(k/\varepsilon)$. Thus, in this case, the algorithm only queries the distances
among points in the sample. Also, they indicated using a simple adversarial construction (of a graph
metric) that $\mathit{\Omega}(\sqrt{n/\varepsilon})$ points are required for the case $\beta < 1$.
For the Euclidean metric ($\ell_2$ metric on $\mathbb{R}^d$) with the radius cost and for the case $\beta = 0$,
is sufficient a sample of size $\widetilde{O}(d k/ \varepsilon)$.
The proof was based on a \emph{VC dimension}\/ argument that, in general, constitutes
a measure of the size of a class of sets.
It is worth mentioning that, the Vapnik-Chervonenkis (VC) dimension was introduced by
Vapnik\footnote{Vladimir Naumovich Vapnik (b. 1936), Uzbek computer scientist.}
and
Chervonenkis\footnote{Alexey Yakovlevich Chervonenkis (1938 -- 2014), Russian mathematician.}
in 1968 \cite{VapnikC1968}, while the corresponding  proofs published in 1971 \cite{VapnikC1971}.
The VC dimension constitutes an important notion in \emph{statistical learning theory}\/ since provides, among others,
a measure of the complexity of a model, related to the quality of its fitness on different data sets
(see,\emph{ e.g}.\ \cite{VovkPG2015}).

Furthermore, the authors of \cite{AlonDPR2004} by using a different analysis for the case $\beta > 0$
proposed that is sufficient a $\widetilde{O}\bigl(k^2 / (\varepsilon \beta^2)\bigr)$-sample.
Also, for the $\ell_2$ metric and the diameter cost, a
$\widetilde{O}\bigl(k^2 / (\varepsilon (2/\beta^2)d\bigr)$-sample suffices.
They also shown, using a high-dimensional geometric argument, that the exponential dependence on $d$ is necessary.
In addition, the authors exhibited how to apply the results to approximate clustering of data.
They pointed out that, their algorithms can also be used to find \emph{approximately good clusterings}.
Specifically, these are clusterings of all but an $\varepsilon$-fraction of the points in $P$ that have optimal
or close to optimal cost.
The advantage of their algorithms is that they create an \emph{implicit representation}\/ of such clusterings
in time independent of ${\rm card}\{P\}=n$.
Therefore, without the necessity of the participation of all the points in $P$,
the implicit representation can be used to answer queries related to the cluster
in which any given point belongs~(\emph{cf}.~\cite{AlonDPR2004}).

\medskip
\noindent
\textbf{Partitioning theorems:}\/
In the directions and trends of property testing a considerable role is played by \emph{partitioning theorems}
that, in general, describe the ways in which convex sets intersect with each other.
Therefore, it is worth mentioning basic significant results related to partitioning theorems,
\emph{i.e}.\ Helly-type theorems~\cite{BaranyK2022}
(see also, \emph{e.g}.\ \cite{Vrahatis2024b} and the references therein).

Carath\'{e}odory's theorem (\emph{cf}.\ Theorem~\ref{CaratheodoryTh}) is related to the following
partitioning theorems due to
Helly\footnote{Eduard Helly (1884 -- 1943), Austrian mathematician.}
and
Tverberg\footnote{Helge Arnulf Tverberg (1935 -- 2020), Norwegian mathematician.}:
\begin{theorem}[Helly's partitioning theorem (1913)~\cite{Helly1923}]\label{HellyTh}
Let $C_1, C_2, \ldots, C_k$ be a finite family of convex subsets of\/ $\mathbb{R}^d$, with $k \geqslant d + 1$.
If the intersection of every $d + 1$ of these sets is nonempty, then the whole family has a nonempty intersection,
\emph{i.e.}~$\cap_{i=1}^{k} C_i \neq \emptyset$.
\end{theorem}

The above theorem proposed by Helly in 1913 \cite{DanzerGK1963}, but it does not published by him until 1923.
\begin{theorem}[Tverberg's partitioning theorem (1966)~\cite{Tverberg1966}]\label{TverbergTh}
Every set with at least $(p - 1)(d + 1) + 1$ points in\/ $\mathbb{R}^d$ can be
partitioned into $p$ subsets whose convex hulls all have at least one point in common.
\end{theorem}

Carath\'{e}odory's, Helly's, and Tverberg's theorems (\emph{cf}.\ Theorems~\ref{CaratheodoryTh}, \ref{HellyTh} and \ref{TverbergTh})
are equivalent in the sense that each one can be deduced from another~\cite{DeLoeraGMM2019}.
Tverberg gave the first proof of Theorem~\ref{TverbergTh} in 1966~\cite{Tverberg1966},
while in 1981 he gave a simpler proof~\cite{Tverberg1981}.
Tverberg's partitioning theorem generalizes the following theorem that it has been proved by
Radon\footnote{Johann Karl August Radon (1887 -- 1956), Austrian mathematician.}
in 1921:
\begin{theorem}[Radon's partitioning theorem (1921)~\cite{Radon1921}]\label{RadonTh}
Every set with $d + 2$ points in $\mathbb{R}^d$ can be partitioned into two sets whose convex hulls intersect.
\end{theorem}

Carath\'{e}odory's, Helly's, Tverberg's, and Radon's theorems
(\emph{cf}.\ Theorems~\ref{CaratheodoryTh}, \ref{HellyTh}, \ref{TverbergTh} and \ref{RadonTh})
 are known as \emph{Helly-type theorems}~\cite{BaranyK2022}.
In addition, a classical partitioning theorem for points in the plane that is based on Carath\'{e}odory's theorem
has been proved by
Birch\footnote{Bryan John Birch (b. 1931), English mathematician.}
in 1959~\cite{Birch1959}.
It is worth mentioning here that,
Adiprasito\footnote{Karim Alexander Adiprasito (b. 1988), German mathematician.},
B\'{a}r\'{a}ny\footnote{Imre B\'{a}r\'{a}ny (b.\ 1947), Hungarian mathematician.},
Mustafa\footnote{Nabil Hassan Mustafa, PhD in 2004 in convex and discrete geometry from the Duke University, USA.}
and
Terpai\footnote{Tam\'{a}s Terpai (b. 1981), Hungarian mathematician and computing scientist.}
in 2020 \cite{AdiprasitoBMT2020} initiated the study of the dimensionless versions
of classical theorems in convexity theory.
Specifically, they considered the dimensionless versions of the theorems of Carath\'{e}odory, Helly, and Tverberg.
The obtained results have several ``colorful'' and ``fractional'' consequences
and are particulary interesting and motivating, among others, for those who are interested in classical convexity
parameters. The authors named these theorems as \emph{``no-dimension theorems''}.

Carath\'{e}odory's theorem has obtained numerous interesting reformulations, variations, generalizations and applications
(see, \emph{e.g}.\
\cite{DanzerGK1963,DeLoeraGMM2019}).
In 1970
Rockafellar\footnotemark[15]
\cite{Rockafellar1970} used Carath\'{e}odory's theorem to prove Helly's theorem and results related to
\emph{infinite systems of linear inequalities}.
In 2018 B\'{a}r\'{a}ny and
Sober\'{o}n\footnote{Pablo Sober\'{o}n (b. 1988), Mexican mathematician.}
\cite{BaranyS2018}
presented an overview related to the advances of Tverberg's theorem.
They discussed, among others, the \emph{topological}, \emph{linear-algebraic}, and \emph{combinatorial}\/
aspects of Tverberg's theorem and its applications.
Also, their survey contains several challenging and interesting open problems and conjectures.
In addition, various generalizations, refinements, applications and conjectures of some of the classical combinatorial
theorems about convex sets (Carathéodory, Helly, Radon and Tverberg theorems) have been published in 2022 by
B\'{a}r\'{a}ny
and
Kalai\footnote{Gil Kalai (b. 1955), Israeli mathematician and computer scientist.}
\cite{BaranyK2022}.
Also, in~\cite{BaranyK2022} the authors point out the connection between important results
from combinatorial convexity and some theorems from topology.
For instance, Helly's theorem   is a manifestation of the \emph{nerve theorem}\/ from algebraic topology,
and Radon’s theorem can be regarded as an early \emph{linear}\/ version of the \emph{Borsuk-Ulam theorem}.

In 1982
B\'{a}r\'{a}ny\footnotemark[44]
\cite{Barany1982} proposed the following
sharp generalization of Carath\'{e}odory's theorem
in the sense that the number of partition of the sets $C_i$'s cannot be decreased.
\begin{theorem}[B\'{a}r\'{a}ny's partitioning theorem (1982)~\cite{Barany1982}]\label{BaranyTh}
Let $C_0, C_1, \ldots, C_d$\, be subsets of\/ $\mathbb{R}^d$
and let $\alpha$ be a point in ${\rm conv}\,C_i$ (the convex hull of $C_i$)
for $i=0,1,\ldots, d$ (\emph{i.e}.\ $\alpha \in \bigcap_{i=0}^d {\rm conv}\,C_i$).
Then there exist vectors ${\upsilon}^i \in C_i$ for $i=0,1,\ldots, d$\,
such that $\alpha \in {\rm conv} \{{\upsilon}^0, {\upsilon}^1, \ldots, {\upsilon}^d \}$.
\end{theorem}

In addition, in a paper published in 1986 by
Polikanova\footnote{Irina Viktorovna Polikanova, PhD in 1987 in convex and discrete geometry from the Novosibirsk State University, Russia.}
and
Perel${}^\prime$man\footnote{Grigorii Yakovlevich Perel${}^\prime\!$man (b.\ 1966), Russian mathematician.}
titled \emph{``A remark on {H}elly's theorem''}, the following partitioning theorem has been proved~\cite{PolikanovaP1986}:
\begin{theorem}[Polikanova-Perel${}^\prime\!$man partitioning theorem (1986)~\cite{PolikanovaP1986}]\label{PolikanovaPTh}
Let $\mathfrak{M}$ be a bounded family of compact convex sets in $\mathbb{R}^d$
such that $\mu_d(\bigcap \mathfrak{M})=0$, where $\mu_d$ denotes the $d$-dimensional volume.
Then for every $\varepsilon > 0$ there are $d+1$ sets
$M_1, M_2, \ldots, M_{d+1}$ in the family $\mathfrak{M}$ such that
$\mu_d \bigl(\bigcap_{i=1}^{d+1} M_i \bigr) < \varepsilon$.
Also, if $\dim \bigcap \mathfrak{M} = \ell$,\, $0\leqslant \ell < d$,
then there are $d-\ell+1$ sets in the family $\mathfrak{M}$
such that $\mu_d \bigl(\bigcap_{i=1}^{d -\ell +1} M_i \bigr) \leqslant \varepsilon$.
\end{theorem}

It is worth emphasizing that, for tackling the clustering problem,
in 2018
Chakraborty\footnote{Sourav Chakraborty, PhD in 2008 in computer science from the University of Chicago, USA.},
Pratap\footnote{Rameshwar Pratap, PhD in 2014 in computer science from the Chennai Mathematical Institute, India.},
Roy\footnote{Sasanka Roy, PhD in 2007 in computer science from the Indian Statistical Institute, Kolkata, India.}
and
Saraf\footnote{Shubhangi Saraf, PhD in 2012 in computer science from the Massachusetts Institute of Technology, USA.}
\cite{ChakrabortyPRS2018}
presented an application of partitioning theorems in the context of property testing.
Specifically, the authors solved the following computational geometry problem, which is equivalent to clustering:
``Given a finite set of points in $\mathbb{R}^d$ and a \emph{proximity parameter},
the following two cases have to be distinguished:
(a) either \emph{all}\/ the points are contained in a unit ball, or
(b) no unit radius ball contains more than a \emph{fraction}\/ of the points
that is expressed in terms of the proximity parameter.''
In addition, the authors presented a simple algorithm in the property testing setting for tackling this problem.
It is worth noting that, the proof of correctness of their algorithm was based on Helly’s partitioning theorem.
Furthermore, their results were generalized by replacing the unit ball by a chosen object
in $\mathbb{R}^d$, together with its translated copies.
Finally, as an application of their testing result, for the case of \emph{clustering with outliers},
the authors have shown that by querying only a constant size of the sample the approximate clusters
can be obtained with high probability.

The authors of \cite{ChakrabortyPRS2018} assumed that the clusters are of a symmetric convex shape.
For a set $P$ of $n$ points in $\mathbb{R}^d$ and a symmetric convex body $S$, they named the set of
points $(k, S)$-\emph{clusterable}\/ if all the points are contained in $k$ translated copies of $S$.
In a promise version of the problem, given a parameter $\varepsilon \in (0, 1]$ the aim
is to distinguish between the cases when $P$ is $(k, S)$-clusterable and when it is
$\varepsilon$-\emph{far from being}\/ $(k, S)$-\emph{clusterable},
\emph{i.e}.\ all $k$ translated copies of $S$ contain at most a $(1-\varepsilon)$ fraction of points.
For the promise version of the problem, they designed randomized algorithm that are generally called \emph{testers}.
Their approach can also be used to find approximately good clusters in the case of clustering with outliers.
They considered the outliers as follows:
``Given a set of $n$ data points, the objective is to remove a set of $z$ points called
outliers such that the radius of a MEB on the remaining $n - z$ points is minimized.''
In this case, they presented a randomized algorithm that takes a constant size sample from the input and outputs
the radius and the center of the cluster.
The advantage of their algorithm is that they construct an approximate representation
of such clustering in running time which is independent of the input size.

The authors presented their result as follows:
\begin{theorem}[Chakraborty-Pratap-Roy-Saraf theorem\! (2018) \cite{ChakrabortyPRS2018}]\label{thm:Chak-etal}
Let $P$ be a set of $n$ points in $\mathbb{R}^d$, $S$ be a symmetric convex body, and
$\varepsilon \in (0, 1]$ be a parameter.
Then, there is an algorithm that samples $O\bigl(\frac{d}{\varepsilon^{d+1}} \bigr)$
many points uniformly at random from the input, and
\begin{itemize}
\leftskip0.05cm
\item[(a)]
accepts, if all the points in $P$ are contained in a translated copy of $S$,
\item[(b)]
 rejects with probability at least $2/3$, if any translated copy of $S$ contains  at most $(1 - \varepsilon) n$ points.
\end{itemize}
The running time of the algorithm is $O\bigl(\frac{d}{\varepsilon^{d+1}} \bigr)$.
\end{theorem}
\vspace{0.1cm}

In addition the authors of \cite{ChakrabortyPRS2018} presented, among others, three algorithms
(see Algorithms \ref{algo:CPRS-1}, \ref{algo:CPRS-2} and \ref{algo:CPRS-3}),
supported by the corresponding theoretical results
(see Theorems \ref{thm:Chak-etal-alg1}, \ref{thm:Chak-etal-alg2} and \ref{thm:Chak-etal-alg3}).
Specifically, the algorithms are related to:
\begin{itemize}
\leftskip0.05cm
\item[(a)]
Algorithm \ref{algo:CPRS-1}: The $(1,S)$-cluster tester for a given symmetric convex body $S$.
\item[(b)]
Algorithm \ref{algo:CPRS-2}: The $(k, G)$-cluster tester for a given geometric objects and
\item[(c)]
Algorithm \ref{algo:CPRS-3}: Approximating clusters with outliers.
\end{itemize}
\begin{algorithm}[ht!]
\textbf{Input:}
A set $P$ of $n$ points in $\mathbb{R}^d$ and $0 < \varepsilon, \delta \leqslant 1$.
\begin{algorithmic}
		\Repeat
\State  select a set (say $W$) of $d + 1$ points uniformly at random
\State  from $P$.
        \If{$W$ is not contained in $S$}
        \State return $W$ as witness.
        \EndIf
		\Until $\frac{1}{\varepsilon^{d+1}} \ln \frac{1}{\delta}$ many times.
        \If{no witness found}
        \State accept
        \Comment {all the points are in a translated copy of $S$.}
        \EndIf
\end{algorithmic}
\textbf{Output:}
Returns a set of $d + 1$ points, if it exists, which are not contained in any translated copy of $S$, or
accepts (\emph{i.e.}\/ all the points are contained in a translated copy of $S$).
\caption{Chakraborty, Pratap, Roy and Saraf (2018) \cite{ChakrabortyPRS2018}:
\emph{The $(1,S)$-cluster tester for a symmetric convex body $S$.}}\label{algo:CPRS-1}
\end{algorithm}
\begin{algorithm}[ht!]
\textbf{Input:}
A set $P$ of $n$ points in $\mathbb{R}^d$, $0 < \delta \leqslant 1$ and $\varepsilon \in (\varepsilon', 1]$.
\begin{algorithmic}
		\Repeat
\State  select a set (say $W$) of $k + 1$ points uniformly at random
\State  from $P$.
        \If{$W$ is not contained in $k$ translated copies of $G$\,}
        \State return $W$ as witness.
        \EndIf
		\Until $\frac{1}{c} \ln \frac{1}{\delta}$ many times.
        \If{no witness found}
        \State accept
        \Comment {all the points are in $k$ translated copies of $G$.}
        \EndIf
\end{algorithmic}
\textbf{Output:}
Returns a set of $k + 1$ points, if it exists, which is not contained in $k$ translated copies of $G$, or
accepts (\emph{i.e.} all the points are contained in it).
\caption{Chakraborty, Pratap, Roy and Saraf (2018) \cite{ChakrabortyPRS2018}:
\emph{The $(k, G)$-cluster tester for geometric objects.}}\label{algo:CPRS-2}
\end{algorithm}
\begin{algorithm}[ht!]
\textbf{Input:}
A set $P$ of $n$ points in $\mathbb{R}^d$ (input is given as black-box) and $0 < \varepsilon, \delta \leqslant 1$.
\begin{algorithmic}[1]
\State Uniformly and independently, select
$m = \frac{d+1}{\varepsilon^{d+1}} \ln \frac{1}{\delta}$
points from $P$.
\State Compute a minimum enclosing ball containing all the sample points and
report its center and radius.
\end{algorithmic}
\textbf{Output:}
Report the center and radius of a cluster which contains all but at most $\varepsilon n$ points.
\caption{Chakraborty, Pratap, Roy and Saraf (2018) \cite{ChakrabortyPRS2018}:
\emph{Approximating clusters with outliers.}}\label{algo:CPRS-3}
\end{algorithm}
\begin{theorem}[Chakraborty-Pratap-Roy-Saraf theorem\!
(2018) \cite{ChakrabortyPRS2018}]\label{thm:Chak-etal-alg1}
Consider a set $P$ of $n$ points in $\mathbb{R}^d$ (where $n \geqslant d+1$)
which are located such that every translated copy of a symmetric convex body $S$ contains less than
$(1 - \varepsilon) n$ points (where $\varepsilon \in (0, 1]$).
Then Algorithm \ref{algo:CPRS-1}, from $\frac{1}{\varepsilon^{d+1}} \ln \frac{1}{\delta}$
(where $\delta \in (0, 1]$) many random samples each of $d + 1$ size sets,
finds a set which is not contained in any translated copy of $S$, with probability at least $1 - \delta$.
\end{theorem}
\begin{theorem}[Chakraborty-Pratap-Roy-Saraf theorem\!
(2018) \cite{ChakrabortyPRS2018}]\label{thm:Chak-etal-alg2}
Consider $k$ translated copies of a geometric object $G$ and a set of $n$ points in $\mathbb{R}^d$
($k$ and $d$ are constants).
Then there exist $\varepsilon{'} = \varepsilon{'}(k, t)$
(where $t$ is a constant that depends on the shape of the geometric object)
such that for all $\varepsilon \in (\varepsilon{'}, 1]$, every $k$ translated copies of $G$
contain at most $(1 - \varepsilon)n$ points.
Algorithm \ref{algo:CPRS-2}, from $\frac{1}{c} \ln \frac{1}{\delta}$ (where $\delta \in (0, 1]$,
$c =\bigl(\frac{1}{2(t+1)(k+1)}\bigr)^{k+1}$, and $c n^{k+1}$ is the number of witnesses)
many random samples each of size $k+1$, finds a set which is not contained in any $k$ translated copies of $G$
with probability at least $1 -\delta$.
\end{theorem}
\begin{theorem}[Chakraborty-Pratap-Roy-Saraf theorem\!
(2018) \cite{ChakrabortyPRS2018}]\label{thm:Chak-etal-alg3}
Given a set of $n$ points in $\mathbb{R}^d$ and $0 < \varepsilon, \delta \leqslant 1$,
Algorithm \ref{algo:CPRS-3} correctly outputs, with probability at least $1 - \delta$,
the center and radius of a ball which contains all but at most $\varepsilon n$ points.
The running time of the algorithm is constant and it queries only a constant size sample
(constant depending on $d$ and $\varepsilon$).
Moreover, if $r_{\rm outlier}$ is the smallest ball containing all but at most $\varepsilon n$ points and
$r_{\min}$ is the smallest ball containing all the points, then Algorithm \ref{algo:CPRS-3} outputs the radius $r$
such that
$r_{\rm outlier} \leqslant r \leqslant r_{\min}$.
\end{theorem}
\vspace*{0.1cm}

The authors of \cite{ChakrabortyPRS2018} have pointed out that, the most interesting part of their result is
the initiation of an application of a Helly-type theorem in property testing for solving the clustering problem.
Also, they proposed a generalization of their considered problem for more than one object.
That is to say, given a natural number $k$ and an object $B$, the aim is to determine with high probability if:
(a) all the $n$ points are contained in $k$ translated copies of $B$, or
(b) any $k$ translated copies of $B$ contains at most a $(1 - \varepsilon)$ fraction of points.
They conjectured that a similar result as of Theorem \ref{thm:Chak-etal} is also valid for the $k$ object translation case.
However, they shown that Helly's theorem does not hold for the $k$ object case.
Nevertheless, they conjectured that a Helly-type theorem such as the Katchalski-Liu fractional version of Helly's theorem,
(see Theorem \ref{thm:KatchalskiL}), does hold in this case.
Also, they pointed out that, if the conjecture is true, then it is possible to obtain a similar result as
of Theorem \ref{thm:Chak-etal} for the $k$ object setting.

In 1979
Katchalski\footnote{Meir Katchalski, PhD in 1974 from the Hebrew University of Jerusalem, Israel.}
and
Liu\footnote{Andrew Chiang-Fung Liu (1947 -- 2024), Canadian mathematician.}
proved the following theorem that it can be considered as a \emph{fractional}\/ version of the Helly theorem:

\begin{theorem}[Katchalski-Liu fractional Helly theorem (1979)
\cite{KatchalskiL1979}]\label{thm:KatchalskiL}
For every $\alpha \in (0, 1]$ there exists $\beta =$ $ \beta(d, \alpha)$  with the following property:
For the convex sets $C_1,$ $ C_2, \ldots, C_n$ in $\mathbb{R}^d$ where $n \geqslant d + 1$,
if at least $\alpha \binom{n}{d+1}$ of the collection of subfamilies of size $d + 1$ has a non-empty intersection,
then there exists a point contained in at least $\beta n$ sets.
\end{theorem}
\vspace{0.1cm}

It is worth noting that, in 1984
Kalai\footnotemark[48]
\cite{Kalai1984} and independently in 1985
Eckhoff\,\footnote{J\"{u}rgen Eckhoff, PhD in 1969 in convex and discrete geometry from the Georg-August-Universit\"{a}t G\"{o}ttingen, Germany.}
\cite{Eckhoff1985}
proved that
$\beta(d, \alpha) = 1 - (1- \alpha)^{\frac{1}{d+1}}$.
In 1985
Alon\footnotemark[34]
and Kalai \cite{AlonK1985}
gave a short proof of this result.
For issues related to Helly-type theorems for \emph{boxes}, the interested reader is referred to \emph{e.g}.\
\cite{Katchalski1980,DanzerG1982,Eckhoff1988,Eckhoff2007,BaranyFMMOP2015,BanosO2018,Dillon2021}.

\section{Enclosing Theorems and the Computation of the Diameter of a Set}\label{sec:Diam}

\medskip
\noindent
\textbf{Enclosing theorems:}\/ It is worth recalling and emphasising that, the MEB problem
can be tackled by applying \emph{enclosing (covering) theorems}.
In 1901
Jung\footnote{Heinrich Wilhelm Ewald Jung (1876 -- 1953), German mathematician.}
gave a seminal enclosing theorem~\cite{Jung1901}.
Specifically, Jung was the first who answered to the question of best possible estimate of the
radius of the smallest spherical surface enclosing a bounded subset $P$ of\, $\mathbb{R}^d$ of
a given diameter, \emph{i.e}.\ the maximal distance of any two points of $P$.
He gave results for the case of {finite point sets} and indicated their extension to {infinite sets},
while in 1910 Jung~\cite{Jung1910} proposed necessary conditions on
the smallest circle enclosing a finite set of~$n$~points in a plane.
It can be considered that, the earliest known interesting application of Jung's theorem was first given in 1905 by
Landau\footnote{Edmund Georg Hermann Landau (1877 -- 1938), German mathematician.}
\cite{Landau1905}
who applied Jung's theorem in a plane in order to sharpen an inequality related to {theory of analytic functions}.
In general, the important Jung enclosing theorem provides an upper bound of the circumradius
of a set in terms of its diameter.
\begin{theorem}[Jung's enclosing theorem (1901)~\cite{Jung1901}]\label{JungTh}
Let ${\rm diam}(P)$ be the diameter of a bounded subset $P$ of\, $\mathbb{R}^d$
(containing more than a single point).
Then, there exists a unique spherical surface of circumradius $\rho_{\rm cir}^d (P)$ enclosing $P$,
and it holds that:
\begin{equation}\label{JungThEq}
\rho_{\rm cir}^d (P) \leqslant \sqrt{\frac{d}{2(d+1)}}\,\, {\rm diam}(P) .
\end{equation}
\end{theorem}

The legacy of Jung’s enclosing theorem remains significant and vibrant, constituting one of the cornerstones
in the field of MEB and related problems that strongly influencing later developments and generalizations
to other spaces and non-Euclidean geometries (see, \emph{e.g}.\ \cite{Vrahatis2024b} and the references therein).

It is worth noticing that, the \emph{Jung Inequality}\/ (\ref{JungThEq}) is \emph{sharp}\/ (best possible) since the
regular $d$-simplex attains equality (\emph{cf}.\ Thm 1 in \cite{Vrahatis2024b}).
However, by considering the \emph{barycentric circumradius}\/ $\beta_{\rm cir}^d=\beta_{\rm cir}^d(P)$
of a set $P$ in $\mathbb{R}^d$ as the supremum of the barycentric circumradii of simplices with vertices in $P$,
it has been shown in \cite{Vrahatis1988}, that a bounded set $P$ can be covered by a spherical surface
of circumradius~$\beta_{\rm cir}^d(P)$, which in many cases gives a better result than Jung's theorem
(\emph{cf.} \cite {Vrahatis2024b,Vrahatis1988}).
\begin{theorem}[Variant of Jung's theorem (1988)~\cite{Vrahatis1988}]\label{VrahatisTheoremScov}
Let ${\rm diam}(P)$ be the diameter of a bounded subset ${P}$ of\, $\mathbb{R}^d$ (containing more than a single point)
and let $\beta_{\rm cir}^d=\beta_{\rm cir}^d(P)$ be its barycentric circumradius.
Then, there exists a unique spherical surface of circumradius $\rho_{\rm cir}^d$ enclosing $P$, and
\begin{equation}\label{eq:pcovrho}
\rho_{\rm cir}^d \leqslant \min \bigl\{\beta_{\rm cir}^d(P),\, [d/(2d+2)]^{1/2}\, {\rm diam}(P) \bigr\}.
\end{equation}
\end{theorem}

For additional enclosing theorems, the interested reader is referred to  \cite{Vrahatis2024b} and the references therein.

It is worth mentioning that, in 2020
Adiprasito\footnotemark[43],
B\'{a}r\'{a}ny\footnotemark[44],
Mustafa\footnotemark[45]
and
Terpai\footnotemark[46]
initiated the study of the dimensionless versions of classical theorems in convexity theory \cite{AdiprasitoBMT2020}.
They proved, among others, the following interesting theorems that provide upper bounds in terms of the diameter of a set:
\begin{theorem}[Adiprasito\,-\,B\'{a}r\'{a}ny\,-\,Mustafa\,-Terpai theorem (2020) \cite{AdiprasitoBMT2020}
-- No-dimension Carath\'{e}odory theorem]\label{NoColCarathTh}
Let $P$ be a set of $n$ points in $\mathbb{R}^d$, $r \in [n]$ (\emph{i.e}.\ $r \leqslant n$), and $a \in {\rm conv}\,\!P$.
Then there exists a subset $Q$ of $P$ with $r$ elements (\emph{i.e}.\ $|Q| = r$) such that for the distance between
$a$ and the convex hull of $Q$ holds that:
\begin{equation}\label{NoColCarathEq}
{\rm dist} \bigl(a,\, {\rm conv}\,Q\bigr) \leqslant \frac{{\rm diam}(P)}{\sqrt{2 r}} .
\end{equation}
\end{theorem}

\begin{theorem}[Adiprasito\,-\,B\'{a}r\'{a}ny\,-\,Mustafa\,-Terpai theorem (2020) \cite{AdiprasitoBMT2020}
-- No-dimension Tverberg theorem]\label{NoColTverbTh}
Let $P$ be a set of $n$ points in $\mathbb{R}^d$, then for a given integer $1 \leqslant k \leqslant n$, there exists
a point $q \in \mathbb{R}^d$ and a partition of $P$ into $k$ sets $P_1, P_2, \ldots, P_k$ such that:
\begin{equation}\label{NoColTverbEq}
{\rm dist} \bigl(q,\, {\rm conv}\,\!P_i\bigr) \leqslant \bigl(2 + \sqrt{2}\bigr)\, \sqrt{ \frac{k}{n}}\,\, {\rm diam}(P),\,\, \forall\, i\in [k].
\end{equation}
\end{theorem}

For additional related theorems, see \emph{e.g}.\ \cite{AdiprasitoBMT2020}.

\medskip
\noindent
\textbf{Computing the diameter of a set:}\/
In 1985
Preparata\footnote{Franco Paolo Preparata (b. 1935), Italian computer scientist.}
and
Shamos\footnote{Michael Ian Shamos (b. 1947), American mathematician.}
gave the following theorem \cite[Thm 4.16]{PreparataS1985}:
\begin{theorem}[Preparata\,-\,Shamos theorem (1985) \cite{PreparataS1985}]\label{PreShaTh}
The computation of the diameter of a finite set of $n$ points in $\mathbb{R}^d$ ($d \geqslant 2$)
requires $\mathit{\Omega}(n \log n)$ operations in the algebraic computation-tree model.
\end{theorem}

Furthermore, the authors pointed out that, for the design of an algorithm for tackling the diameter problem,
the main issue for possibly avoiding the examination of all pairs of points can be provided by the following lemma due to
Hocking\footnote{John Gilbert Hocking (1920 -- 2011), American mathematician.}
and
Young\footnote{Gail Sellers Young, Jr. (1911 -- 1999), American mathematician.}
\cite[Lem 5-17]{HockingY1961}:
\begin{lemma}[Hocking\,-\,Young lemma (1961) \cite{HockingY1961}]\label{HocYouLem}
The diameter of the convex hull of any set $A$ is equal to the diameter of $A$ itself.
\end{lemma}

Also, in \cite{PreparataS1985} is emphasized that, in the worst case, all of the original points of the set
could be vertices of the hull, thus it is required $\mathit{O}(n \log n)$ time without eliminating anything.

In the context of the maximum distance between $n$ points, the following two classical theorems that have been proved by
Erd\"{o}s\footnote{Paul (P\'{a}l) Erd\"{o}s (1913 -- 1996), Hungarian mathematician.}
in \cite{Erdos1946} and  \cite{Erdos1960} are presented (\emph{cf}.\ \cite{PreparataS1985}):
\begin{theorem}[Erd\"{o}s theorem (1946) \cite{Erdos1946}]\label{ErdTh1946}
Let the maximum and minimum distances determined by $n$ points in a plane be denoted by $r$ and $r'$, respectively.
Then $r$  can occur at most $n$ times and $r'$  at most $3 n - 6$ times.
\end{theorem}
\begin{theorem}[Erd\"{o}s theorem (1960) \cite{Erdos1960}]\label{ErdTh1960}
The maximum distance between $n$ points in $d$-space can
occur $n^2 / (2 - 2 / \lfloor d/2 \rfloor) +
\mathit{O}(n^{2-\varepsilon})$ times, for $d > 3$ and some $\varepsilon > 0$.
\end{theorem}

In general, using the brute-force algorithm that compares the distances between all pairs of points,
the $\mathit{O}(n^2)$ upper-bound can be obtained.
Better bounds can be obtained in the case of $d=2$ and $d=3$.
Specifically, in the case of $d=2$ the problem can optimally be solved in $\mathit{O}(n \log n)$.
In the case of $d=3$ a randomized $\mathit{O}(n \log n)$ algorithm has been given in 1989 by
Clarkson\footnotemark[21]
and
Shor\footnote{Peter Williston Shor (b. 1959), American mathematician.}
\cite{ClarksonS1989}.
This algorithm requires the computation of the intersection of $n$ balls of the same radius in $\mathbb{R}^3$
and  with respect to this intersection, it requires the rapid localization of points.
Thus, for almost any data set this algorithm exhibits a less efficient behavior than the brute-force algorithm.
Moreover, this algorithm in higher dimensional cases is not efficient, since the intersection
of $n$ balls of the same radius requires $\mathit{\Theta}\bigl(n^{\lfloor d/2 \rfloor}\bigr)$.
In general, the three-dimensional diameter problem can be solved in $\mathit{O}(n \log^3 n)$
as it has been proposed in 1997 by
Ramos\footnote{Edgar Arturo Ramos, PhD in 1995 in convex and discrete geometry from the University of Illinois at Urbana-Champaign, USA.}
\cite{Ramos1997}
and it can be tackled with deterministic algorithms in $\mathit{O}(n \log^2 n)$ as it has been given in 1998 by
Bespamyatnikh\footnote{Sergei Bespamyatnikh, University of British Columbia, Canada.}
\cite{Bespamyatnikh1998}.
Also, an optimal $\mathit{O}(n \log n)$ deterministic algorithm has been proposed in 2000 by Ramos~\cite{Ramos2000}.
The preceding approaches require complicated data structures and techniques including, among others, convex hulls in $\mathbb{R}^3$,
farthest-point
Voronoi\footnote{Georgy Feodosevich Voronoi (or Vorono\"{i}, Voronoy, Voronyi) (1868 -- 1908), Ukrainian mathematician.}
diagrams, intersection of balls and point location search structures.

A possible extension of the above methods to higher dimensions is not attractive.
This is so because the data structures that they apply have sizes that depend exponentially on the dimension,
\emph{e.g}.\ the convex hull of $n$ points in $\mathbb{R}^d$ requires $\mathit{\Omega}\bigl(n^{\lfloor d/2 \rfloor}\bigr)$.
An algorithm that does not require the computation of convex hulls and it can be applied in any dimension
has been proposed in 2002 by
Malandain\footnote{Gr{\'e}goire Malandain, PhD in 1992 in medical image analysis from the {\'E}cole Centrale des Arts et Manufactures, France.}
and
Boissonnat\footnote{Jean-Daniel Boissonnat (b. 1953), French computer scientist.}
\cite{MalandainB2002-a,MalandainB2002-b,MalandainB2002-c}.
It does not require the points to be in a convex position and the only numerical computation burden is the dot product computations
similar to the brute-force algorithm.
Although the algorithm is not worst-case optimal, it appears to be very rapid for a large variety of point distributions.
An exception occurring when the points are distributed on a domain of constant width, \emph{e.g}.\ on a sphere.
A similar to the above algorithm has independently been proposed in 2001 by
Har-Peled\footnote{Sariel Har-Peled (b. 1971), Israeli-American computer scientist.}
\cite{Har-Peled2001}.
The preceding algorithms can be combined together in order to obtain advantage of both as it has been pointed out
by Malandain and Boissonnat in 2002~\cite{MalandainB2002-a}.

In many applications that are related to large data sets the input can be viewed as a \emph{data stream}\/
that the applied algorithm reads in one pass (see, \emph{e.g}.\ \cite{FeigenbaumKSV2002}).
In a \emph{streaming model}, the complexity of an algorithm is measured by the number of passes it makes
over the input data, and the amount of memory used over the execution of the algorithm.
In some cases the input stream is infinite, and the application uses only recent data.
In this case the \emph{sliding-window model}\/ could be more appropriate.
Also, a sliding-window algorithm should run through the input stream once.
In addition, there is not enough storage space for all the data, even for the data in one window
(see, \emph{e.g}.\ \cite{DatarGIM2002}).

A related approach, named \emph{$k$-windows algorithm}\/ has been introduced in 2002
for improving the well known and widely used \emph{$k$-means}\/ clustering algorithm
\cite{VrahatisBAP2002}.
The $k$-windows is an unsupervised algorithm, which by employing windowing techniques, attempts not only to
discover the clusters but also their number, in a single execution.
Also, it can be easily parallelized and it can track the evolution of cluster models in dynamically changing databases,
without a significant computational overhead.
For details and applications, the interested reader is referred to \emph{e.g}.\
 \cite{AlevizosTV2004,BoutsinasTV2006,TasoulisV2005,TasoulisV2007,AntzoulatosV2011}.

In 2005
Feigenbaum\footnote{Joan Feigenbaum (b. 1958), American computer scientist.},
Kannan\footnote{Sampath Kumar Kannan, PhD in 1990 in computer science from the University of California, Berkeley, USA.}
and
Zhang\footnote{Jian Zhang, PhD in 2005 from the Yale University, USA.}
\cite{FeigenbaumKZ2005}
studied the \emph{diameter problem}\/ in the \emph{streaming}\/ and \emph{sliding-window models}.
They shown that, for a stream of $n$ points or a sliding window of size $n$, any exact algorithm for diameter requires $\varOmega(n)$
bits of space.
By assuming that $A$ denotes the output of an algorithm and $T$ the value of the function that the algorithm wants to
compute, then the authors defined that, $A$ $\varepsilon$-\emph{approximates}\/ $T$ if $1 > \varepsilon > 0$ and
$(1 + \varepsilon) T \geqslant A \geqslant (1 - \varepsilon) T$.
Also, they  presented a simple \emph{$\varepsilon$-approximation algorithm}\/ for computing the diameter in the streaming model.
Their main result is an $\varepsilon$-approximation algorithm that maintains the diameter in two dimensions in the sliding-window model using
$O\bigl((1/\varepsilon^{3/2}) \log^3 n\,(\log R + \log \log n + \log(1/\varepsilon))\bigr)$ bits of space, where $R$ is the
maximum, over all windows, of the ratio of the diameter to the minimum non-zero distance between any two points in the window.

In 2022
Pellizzoni\footnote{Paolo Pellizzoni, University of Padova, Padova, Italy.},
Pietracaprina\footnote{Andrea Pietracaprina, University of Padova, Padova, Italy.}
and
Pucci\footnote{Geppino Pucci, University of Padova, Padova, Italy.}
\cite{PellizzoniPP2022}
presented novel \emph{streaming algorithms}\/ for the \emph{$k$-center}\/ and the \emph{diameter}\/ estimation problems
for general metric spaces in the context of the \emph{sliding-window model}.
The central concept used for their algorithms was the maintenance of a small core-set.
This core-set, at any time, allows to compute a solution to the problem under consideration for the current window.
Its quality can be arbitrarily close to the one of the best solution attainable by running a polynomial-time
sequential algorithm on the entire window.
Note that, the size of their core-sets is independent of the window length.
It can be upper bounded by a function of the target number of centers (for the $k$-center problem), of the desired accuracy,
and of the characteristics of the current window, namely its doubling dimension and aspect ratio.
Also, one of the major strengths of their algorithms is that they adapt obliviously to these two latter characteristics.

\section{Concluding Remarks}\label{sec:Synop}

Methodology towards the solution of the MEB problem has been provided.
Mathematical formulation and typical methods for solving this problem have been presented.

A characteristic for the solution of the MEB problem is that a large variety of methods and algorithms
have been proposed in the literature.
Thus, the selection of the approaches that have been presented in the paper at hand,
is neither exhaustive nor strictly representative.
Therefore, the paper has been restricted to the following interesting research areas  that are
related to the MEB problem, namely:
(a) promise problems and property testing,
(b) theorems for enclosing, covering and partitioning a set,
and (c) computation of the diameter of a set.

We hopefully think that the paper at hand might contribute to a rigorous tackling of the MEB problem and
the similar and related to it issues and aspects.

\vfill
\begin{IEEEbiography}[{\includegraphics[width=1in,height=1.25in,clip,keepaspectratio]{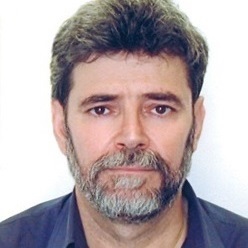}}]{Michael N. Vrahatis}
is a professor emeritus of the University of Patras, Greece.
His main scientific objectives are:
(a) the investigation and mathematical foundation of methods in artificial intelligence,
(b)~the development of innovative methods in this field,
(c) the application of the acquired expertise to address challenging real-life problems in different branches of science and
the search of directions of theoretical mathematical results towards applications, and
(d) the dissemination of knowledge
to young researchers and scholars.
His current interests focus on the thematic areas:
(a) mathematics,
(b) natural computing and computational intelligence,
(c) global optimization,
(d) reliable computing and imprecise data, and
(e) neural networks and machine learning~\& intelligence.
\end{IEEEbiography}
\end{document}